\documentclass[11pt]{article}

\usepackage{amsmath}
\usepackage{amsthm}
\usepackage{amssymb}
\usepackage{amscd}

\theoremstyle{definition}
\newtheorem{theorem}{Theorem}[section]
\newtheorem{prop}[theorem]{Proposition}
\newtheorem{lemma}[theorem]{Lemma}
\newtheorem{corollary}[theorem]{Corollary}

\newtheorem{remark}[theorem]{Remark}

\numberwithin{equation}{section}
\newenvironment{demo}[1]{%
  \trivlist
  \item[\hskip\labelsep
        {\it #1.}]
}{%
\hfill\qedsymbol
  \endtrivlist
}

\newcommand\Int{\mathbb{Z}}

\newcommand\Comp{\mathbb{C}}

\newcommand\Pf{\operatorname{Pf}}
\newcommand\sgn{\operatorname{sgn}}
\newcommand\Sym{\mathfrak{S}} 

\renewcommand\hat{\widehat}
\renewcommand\tilde{\widetilde}
\newcommand\be{\boldsymbol{e}}
\newcommand\bv{\boldsymbol{v}}
\newcommand\trans{{}^t\!}

\newcommand\inv{\operatorname{inv}}

\newcommand\UGM{\operatorname{UGM}}

\allowdisplaybreaks

\title{
On the Expansion Coefficients of KP Tau Function\\
}

\author{
Atsushi Nakayashiki
\footnote{
Department of Mathematics, Tsuda University,
Kodaira, Tokyo 187-8577, Japan,
{\tt atsushi@tsuda.ac.jp}
}
\footnote{
This work was partially supported by the JSPS Grants-in-Aid for Scientific Research No. 15K04907.
},
\ 
Soichi Okada
\footnote{
Graduate School of Mathematics, Nagoya University, 
Furo-cho, Chikusa-ku, Nagoya 464-8602, Japan, 
{\tt okada@math.nagoya-u.ac.jp}
}
\footnote{
This work was partially supported by the JSPS Grants-in-Aid for Scientific Research No. 15K13425.
},
\ 
and 
Yoko Shigyo
\footnote{
Department of Mathematics, Tsuda University,
Kodaira, Tokyo 187-8577, Japan,
{\tt y.shigyo@tsuda.ac.jp}
}
}

\date{
}

\begin{document}

\maketitle

\centerline{\it Dedicated to Masaki Kashiwara on his 70th birthday.}

\begin{abstract}
We study the expansion coefficients of the tau function of the KP hierarchy.
If  the tau function does not vanish at the origin, 
it is known that the coefficients are given by Giambelli formula 
and that it characterizes solutions of the KP hierarchy.
In this paper, we find a generalization of Giambelli formula 
to the case when the tau function vanishes at the origin. 
Again it characterizes solutions of the KP hierarchy.
\end{abstract}

\section{%
Introduction
}

As discovered by Sato \cite{SS}, the $\tau$-function of the KP hierarchy can be expanded by 
Schur functions $s_\lambda(x)$ in the so-called Sato variables $x = (x_1, x_2, \dots)$ 
as
\begin{equation}
\label{eq:tau}
\tau(x) = \sum_\lambda \xi_\lambda s_\lambda(x),
\end{equation}
and the KP hierarchy, which is a system of nonlinear differential equations, reduces 
to Pl\"ucker relations for the coefficients $\{ \xi_\lambda \}_\lambda$.
This is the reason why solutions of the KP hierarchy are parametrized 
by an infinite dimensional Grassmannian, which we call the Sato Grassmannian, and 
$\xi_\lambda$ becomes the Pl\"ucker coordinate of a point of it.
From this description another characterization of the solution of the KP hierarchy 
in terms of coefficients is obtained \cite{EH}.
Namely, if $\xi_\emptyset =1$, then (\ref{eq:tau}) is a solution of the KP hierarchy 
if and only if the coefficients $\{ \xi_\lambda \}_\lambda$ satisfy the following Giambelli formula:
\begin{equation}
\label{eq:KP_Giambelli}
\xi_{(\alpha_1, \dots, \alpha_r|\beta_1, \dots, \beta_r)}
 =
\det \left( \xi_{(\alpha_i|\beta_j)} \right)_{1 \le i, j \le r},
\end{equation}
where $(\alpha_1, \dots, \alpha_r|\beta_1, \dots, \beta_r)$ is the Frobenius notation of a partition.
Notice that, if  $\xi_\emptyset \neq 0$ then it is always possible to normalize $\xi_\emptyset =1$ by 
multiplying $\tau(x)$ by $\xi_\emptyset^{-1}$. 
Recently formulas like (\ref{eq:KP_Giambelli}) play an important role in connecting quantum integrable systems 
to classical integrable hierarchies \cite{AKLTZ,ALTZ}.
They also have an application to the study of higher genus theta functions \cite{EEG}.
However, in the latter case, it is indispensable to consider the case of $\xi_\emptyset=0$ \cite{N1,EEG,N2}.
In this paper we study a generalization of (\ref{eq:KP_Giambelli}) to the case of $\xi_\emptyset = 0$.

Using the Sato Grassmannian it is possible to express $\xi_\lambda$ as a determinant.
However, even in the case $\xi_\emptyset = 1$, this determinant formula is not identical 
to (\ref{eq:KP_Giambelli}).
We have to transform the Grassmann determinant formula in some way.
From this point of view the problem is to formulate a reasonable formula which can be regarded 
as a natural generalization of (\ref{eq:KP_Giambelli}).

Originally the Giambelli identity \cite{G} (see \cite[I.3. Example~9]{Mac}) is the formula for Schur functions:
\begin{equation}
\label{eq:Schur_Giambelli}
s_{(\alpha_1, \dots, \alpha_r|\beta_1, \dots, \beta_r)}(x)
 =
\det \left( s_{(\alpha_i|\beta_j)}(x) \right)_{1 \le i, j \le r}.
\end{equation}
It can be shown that, for any fixed partition $\mu$, 
skew Schur functions $\{ s_{\lambda/\mu} \}_\lambda$ become Pl\"ucker coordinates of the KP hierarchy.
So from the view point of symmetric functions, our problem is related with a generalization 
of the Giambelli identity to skew Schur functions.

The hint to solve the problem comes from the study of the BKP hierarchy, 
where the $\tau$-function is expanded by Schur's $Q$-functions
$$
\tau(x) = \sum_\lambda \zeta_\lambda Q_\lambda \left( \frac{x}{2} \right),
$$
where $\lambda$ runs over all strict partitions.
In the case of $\zeta_\emptyset = 1$, 
a similar formula to (\ref{eq:KP_Giambelli}) is known also for the BKP hierarchy \cite{DJKM1}.
One of the authors of the present paper (Y.S.) has generalized it 
to the case of $\zeta_\emptyset=0$ completely \cite{Shig2}.
The method is very suggestive.
The formula for $\zeta_\lambda$ is derived by solving the equations for $\{ \zeta_\lambda \}_\lambda$, 
which is obtained by expanding the addition formula for $\tau$-function \cite{Shig1}. 
In the KP case the equations obtained from the addition formula of $\tau$-function \cite{SS} 
are nothing but the Pl\"ucker relations.
However it is difficult to do a similar calculation in the usual form of Pl\"ucker relations.
We have found that, if we rewrite the labels of Pl\"ucker relations in terms of Frobenius notation 
of a partition, then it is possible to rewrite Pl\"ucker relations in a similar form 
to the relations for the BKP hierarchy given in \cite{Shig2}.
Then we find a natural generalization of (\ref{eq:KP_Giambelli}) as follows.

\begin{theorem}
\label{thm:main}
For a fixed partition $\mu = (\gamma_1, \dots, \gamma_s|\delta_1, \dots, \delta_s)$, 
the following are equivalent:
\begin{enumerate}
\item[(i)]
The function $\tau(x)$ of the form (\ref{eq:tau}) is a solution of the KP hierarchy 
satisfying
\begin{equation}
\label{eq:cond1}
\xi_\lambda = 0
\quad\text{if $|\lambda| < |\mu|$ or if $|\lambda| = |\mu|$ and $\lambda \neq \mu$,}
\end{equation}
and
\begin{equation}
\label{eq:cond2}
\xi_\mu = 1.
\end{equation}
\item[(ii)]
For any partition $\lambda = (\alpha_1, \dots, \alpha_r | \beta_1, \dots, \beta_r)$, 
the corresponding coefficient $\xi_\lambda$ is expressed as
\begin{equation}
\label{eq:det}
\xi_\lambda
 =
(-1)^s
\det
\begin{pmatrix}
 \big( z_{\alpha_i, \beta_j} \big)_{1 \le i, j \le r}
&
 \big( u_{\alpha_i}^{(j)} \big)_{1 \le i \le r, 1 \le j \le s}
\\
 \big( v_{\beta_j}^{(i)} \big)_{1 \le i \le s, 1 \le j \le r}
&
 O
\end{pmatrix},
\end{equation}
where $z_{a,b}$, $u_a^{(j)}$, $v_b^{(i)}$ satisfy the following conditions:
\begin{equation}
\label{eq:entry}
\left\{
\begin{aligned}
z_{a,b} &= 0
\quad\text{if $a \in \{ \gamma_1, \dots, \gamma_s \}$ or $b \in \{ \delta_1, \dots, \delta_s \}$,}
\\
u_a^{(j)} &= 0
\quad\text{if $a \in \{ \gamma_1, \dots, \widehat{\gamma_j}, \dots, \gamma_s \}$ or $a < \gamma_j$,}
\\
u_{\gamma_j}^{(j)} &= (-1)^{j-1},
\\
v_b^{(i)} &= 0
\quad\text{if $b \in \{ \delta_1, \dots, \widehat{\delta_i}, \dots, \delta_s \}$ or $b < \delta_i$,}
\\
v_{\delta_i}^{(i)} &= (-1)^{i-1}.
\end{aligned}
\right.
\end{equation}
\end{enumerate}
\end{theorem}

We note that the entries of the determinant in (\ref{eq:det}) are given by
\begin{align*}
z_{a,b}
 &=
\xi \begin{pmatrix} a, \gamma_1, \dots, \gamma_s \\ b, \delta_1, \dots, \delta_s \end{pmatrix},
\\
u_a^{(j)}
 &=
\xi \begin{pmatrix} a, \gamma_1, \dots, \widehat{\gamma_j}, \dots, \gamma_s \\ \delta_1, \dots, \delta_s \end{pmatrix},
\\
v_b^{(i)}
 &=
\xi \begin{pmatrix} \gamma_1, \dots, \gamma_s \\ b, \delta_1, \dots, \widehat{\delta_i}, \dots, \delta_s \end{pmatrix}.
\end{align*}
See Proposition~\ref{prop:entry}.

As a corollary of Thoerem~\ref{thm:main}, we obtain a generalization of the Giambelli identity 
(\ref{eq:Schur_Giambelli}) for skew Schur functions (Corollary~\ref{cor:Schur_skewGiambelli}).

The present paper is organized as follows.
In Section~2 we recall the KP hierarchy and the corresponding Pl\"ucker relations, 
and rephrase the Pl\"ucker relations in terms of Frobenius notation of partitions.
Section~3 is devoted to the proof of Theorem~\ref{thm:main}.
In Section~4 we compute the Pl\"ucker coordinates of the frame of a point of the Sato Grassmannian 
assuming a generic condition and show that the
determinant formula in Theorem~\ref{thm:main} can be derived in this way.
In Section~5, we apply our main theorem to derive a generalization of the Giambelli identity to 
skew Schur functions.

\section{%
KP hierarchy
}

In this section, we review the KP hierarchy in the bilinear form 
and rephrase the corresponding Pl\"ucker relations in terms of Frobenius notation.

The KP hierarchy in the bilinear residue form \cite{DJKM2} is the equation for the function $\tau(x)$ of 
$x = (x_1, x_2, \dots)$ given by
\begin{equation}
\int \tau(x-y-[k^{-1}]) \tau(x+y+[k^{-1}]) \exp \left( -2 \sum_{j=1}^\infty y_j k^j \right) dk
 =
0,
\label{eq:KP}
\end{equation}
where $[k^{-1}] = (k^{-1}, k^{-1}/2, k^{-3}/3, \dots)$, $y  =(y_1, y_2, \dots)$ 
and the integral denotes taking the residue at $k = \infty$, i.e. the coefficient of $k^{-1}$ 
in the Laurent expansion.
If we expand this equation in the variable $y$, 
then we get an infinite set of bilinear differential equations for $\tau(x)$. 

We consider formal power series solutions of the KP hierarchy. 
Since any formal power series in $x$ can be expanded by Schur functions, 
we can write $\tau(x)$ as 
$$
\tau(x)
 =
\sum_{\lambda} \xi_\lambda s_\lambda(x),
$$
where $\lambda$ runs over all partitions.
The KP hierarchy (\ref{eq:KP}) reduces to the Pl\"ucker relations for the coefficients $\{ \xi_\lambda \}$. 

A subset $M \subset \Int$ is called a Maya diagram of charge $c$ 
if both $\Int_{\ge 0} \cap M$ and $\Int_{<0} \setminus M$ are finite and 
$\# (\Int_{\ge 0} \cap M) - \# (\Int_{<0} \setminus M) = c$, 
where $\Int_{\ge 0}$ (resp $\Int_{<0}$) denotes the set of all nonnegative (resp. negative) integers.
We often represent a Maya diagram $M$ as the decreasing sequence of elements in $M$.

A partition is a weakly decreasing sequence $\lambda = (\lambda_1, \lambda_2, \lambda_3, \dots)$ 
of nonnegative integers such that $|\lambda| = \sum_{i \ge 1} \lambda_i$ is finite.
We identify a partition $\lambda$ with its Young diagram, 
which is a left-justified array of $|\lambda|$ cells with $\lambda_i$ cells in the $i$th row.
Given a partition $\lambda$, we put
$$
p(\lambda) = \# \{ i : \lambda_i \ge i \},
\quad
\alpha_i = \lambda_i - i,
\quad
\beta_i = \lambda'_i - i
\quad(1 \le i \le p(\lambda)),
$$
where $\lambda'_i$ is the number of cells in the $i$th column of the Young diagram of $\lambda$.
Then we write $\lambda = (\alpha_1, \dots, \alpha_{p(\lambda)} | \beta_1, \dots, \beta_{p(\lambda)})$ 
and call it the Frobenius notation of $\lambda$.

Maya diagrams $M$ of charge $0$ are in bijection with partitions $\lambda$ via
$$
M = (\lambda_1 - 1, \lambda_2 - 2, \lambda_3 - 3, \dots),
$$
and
$$
\Int_{\ge 0} \cap M = \{ \alpha_1, \dots, \alpha_r \},
\quad
\Int_{<0} \setminus M = \{ - \beta_1 - 1, \dots, - \beta_r -1 \},
$$
where $\lambda = (\alpha_1, \dots, \alpha_r | \beta_1, \dots \beta_r)$ in the Frobenius notation.

If a partition $\lambda = (\alpha_1, \dots, \alpha_r | \beta_1, \dots, \beta_r)$ 
corresponds to a Maya diagram $M = ( m_1, m_2, \dots )$, 
we write
$$
\xi_\lambda
 =
 \xi \begin{pmatrix} \alpha_1, \dots, \alpha_r \\ \beta_1, \dots, \beta_r \end{pmatrix}
 =
 \xi[m_1, m_2, \dots]
$$
for the coefficient of the Schur function $s_\lambda(x)$ in (\ref{eq:tau}).
Hence we have
\begin{align}
\tau(x)
 &=
\sum_{\alpha, \beta}
 \xi \begin{pmatrix} \alpha_1, \dots, \alpha_r \\ \beta_1, \dots, \beta_r \end{pmatrix}
 s_{(\alpha_1, \dots, \alpha_r | \beta_1, \dots, \beta_r)} (x)
\label{eq:tauF}
\\
 &=
\sum_M \xi[m_1, m_2, \dots] s_{(m_1+1,m_2+2, \dots)}(x),
\label{eq:tauM}
\end{align}
where $(\alpha, \beta)$ runs over all pairs of decreasing sequences 
$\alpha = (\alpha_1, \dots, \alpha_r)$ and $\beta = (\beta_1, \dots, \beta_r)$ 
of nonnegative integers of the same length, 
and $M = (m_1, m_2, \dots)$ runs over all Maya diagrams of charge $0$.
We extend the definition of $\xi \begin{pmatrix} \alpha_1, \dots, \alpha_r \\ \beta_1, \dots, \beta_r \end{pmatrix}$ 
to any pair of sequences of nonnegative integers of the same length 
so that $\xi \begin{pmatrix} \alpha_1, \dots, \alpha_r \\ \beta_1, \dots, \beta_r \end{pmatrix}$ is 
skew-symmetric in $\alpha_1, \dots, \alpha_r$ and $\beta_1, \dots, \beta_r$ respectively.
Also we extend the definition of $\xi[m_1, m_2, \dots ]$ to any integer sequences 
so that $\xi[ m_1, m_2, \dots ]$ is skew-symmetric in $m_1, m_2, \dots$.

\begin{prop}\cite{SS}
\label{prop:KP_Plucker}
The function $\tau(x)$ given as (\ref{eq:tau}) is a solution of the KP hierarchy 
if and only if the coefficients $\{ \xi[M] \}_M$ satisfy the Pl\"ucker relations
\begin{equation}
\label{eq:KP_Plucker}
\sum_{i \ge 1}
 (-1)^i
 \xi[m_1, m_2, \dots, \hat{m_i}, \dots]
 \xi[m_i, n_1, n_2, \dots]
 =
0,
\end{equation}
for any Maya diagrams $M = (m_1, m_2, \dots)$ and $N = (n_1, n_2, \dots)$ of charge $1$ and $-1$ respectively,
where the symbol $\hat{m}$ means removing $m$ from the sequence.
\end{prop}

The following form of the Pl\"ucker relations is the key to our argument.

\begin{prop}
\label{prop:KP_PluckerF}
The function $\tau(x)$ is a solution of the KP hierarchy if and only if the coefficients $\xi_\lambda$ satisfy 
the following Pl\"ucker relations:
\begin{multline}
\label{eq:KP_PluckerF}
\sum_{i=1}^{p+1}
 (-1)^i
 \xi \begin{pmatrix} m_1, \dots, \widehat{m_i}, \dots, m_{p+1} \\ m'_1, \dots, m'_p \end{pmatrix}
 \xi \begin{pmatrix} m_i, n_1, \dots, n_q \\ n'_1, \dots, n'_{q+1} \end{pmatrix}
\\
 =
\sum_{j=1}^{q+1}
 (-1)^{p+j}
 \xi \begin{pmatrix} m_1, \dots, m_{p+1} \\ m'_1, \dots, m'_p, n'_j \end{pmatrix}
 \xi \begin{pmatrix} n_1, \dots, n_q \\ n'_1, \dots, \widehat{n'_j}, \dots, n'_{q+1} \end{pmatrix},
\end{multline}
for any sequences $m_1, \dots, m_{p+1}, m'_1, \dots, m'_p, n_1, \dots, n_q, n'_1, \dots, n'_{q+1}$ 
of nonnegative integers.
\end{prop}

\begin{demo}{Proof}
First note that the both sides of (\ref{eq:KP_PluckerF}) are alternating in 
$(m_1, \dots, m_{p+1})$, $(m'_1, \dots, m'_p)$, $(n_1, \dots, n_q)$ and $(n'_1, \dots, n'_{q+1})$ respectively.
This fact follows from the following relation between inversion numbers
$$
\inv( i_1, \dots, \hat{i_k}, \dots, i_n )
 \equiv
\inv( i_1, \dots, i_n) - k + i_k
\ \bmod 2,
$$
where $(i_1, \dots, i_n)$ is a sequence of distinct positive integers, 
and $\inv (a_1, \dots, a_m)$ is the number of pairs $(i,j)$ such that $i < j$ and $a_i > a_j$.
Hence we may assume that $m_1 > \dots > m_{p+1}$, $m'_1 > \dots > m'_p$, 
$n_1 > \dots > n_q$ and $n'_1 > \dots > n'_{q+1}$ in (\ref{eq:KP_PluckerF}).

Let $M$ and $N$ be Maya diagrams of charge $1$ and $-1$ respectively 
such that
\begin{gather*}
\Int_{\ge 0} \cap M = \{ m_1 > \dots > m_{p+1} \},
\quad
\Int_{<0} \setminus M = \{ - m'_p -1 > \dots > -m'_1 - 1 \},
\\
\Int_{\ge 0} \cap N = \{ n_1 > \dots > n_q \},
\quad
\Int_{<0} \setminus N = \{ - n'_{q+1}-1 > \dots > -n'_1 - 1 \}.
\end{gather*}
Let $m_i$ be the $i$th largest element of $M$.
If $1 \le i \le p+1$, then $m_i \ge 0$ and we have
\begin{gather*}
\xi[m_1, \dots, \widehat{m_i}, \dots]
 =
\xi \begin{pmatrix} m_1, \dots, \widehat{m_i}, \dots, m_{p+1} \\ m'_1, \dots, m'_p \end{pmatrix},
\\
\xi[m_i, n_1, n_2, \dots]
 =
\xi \begin{pmatrix} m_i, n_1, \dots, n_q \\ n'_1, \dots, n'_{q+1} \end{pmatrix}.
\end{gather*}
Suppose $i \ge p+2$. In this case $m_i < 0$.
If $m_i \in N$, then we have
$$
\xi[m_i, n_1, n_2, \dots] = 0.
$$
If $m_i \not\in N$, then there exists an index $j$ such that $m_i = - n'_j - 1$.
Let $k$ be the largest integer satisfying $m'_k > n'_j$ (if $n'_j > m'_1$ then we put $k=0$).
Then we see that
$$
i = \# \{ m \in M : m \ge -n'_j - 1 \} = (p+1) + (n'_j+1) - (p-k) = n'_j + k+2,
$$
and
\begin{align*}
\xi[ m_1, \dots, \widehat{m_i}, \dots]
 &=
\xi \begin{pmatrix}
 m_1, \dots, m_{p+1} \\ m'_1, \dots, m'_k, n'_j, m'_{k+1}, \dots, m'_p
\end{pmatrix}
\\
 &=
(-1)^{p-k}
\xi \begin{pmatrix}
 m_1, \dots, m_{p+1} \\ m'_1, \dots, m'_k, m'_{k+1}, \dots, m'_p, n'_j
\end{pmatrix}.
\end{align*}
Since the number of elements $n \in N$ greater than $m_i = -n'_j-1$ is equal to 
$q+n'_j-(q+1-j) =n'_j+j-1$, we have
$$
\xi[m_i, n_1, n_2, \dots]
 =
(-1)^{n'_j+j-1}
\xi \begin{pmatrix}
 n_1, \dots, n_q \\
 n'_1, \dots, \hat{n'_j}, \dots, n'_{q+1}
\end{pmatrix}.
$$
Hence we have
\begin{align*}
&
\sum_{i \ge 1}
 (-1)^i
 \xi[ m_1, \dots, \widehat{m_i}, \dots ]
 \xi[ m_i, n_1, n_2, \dots ]
\\
&\quad
=
\sum_{i=1}^{p+1}
 (-1)^i
 \xi \begin{pmatrix}
  m_1, \dots, \widehat{m_i}, \dots, m_{p+1} \\
  m'_1, \dots, m'_p
 \end{pmatrix}
 \xi \begin{pmatrix}
  m_i, n_1, \dots, n_q \\
  n'_1, \dots, n'_{q+1}
 \end{pmatrix}
\\
&\quad\quad
+
\sum_{j=1}^{q+1}
 (-1)^{p+j-1}
 \xi \begin{pmatrix}
  m_1, \dots, m_{p+1} \\
  m'_1, \dots, m'_p, n'_j
 \end{pmatrix}
 \xi \begin{pmatrix}
  n_1, \dots, n_q \\
  n'_1, \dots, \hat{n'_j}, \dots, n'_{q+1}
 \end{pmatrix}.
\end{align*}
Therefore we can conclude that the Pl\"ucker relations (\ref{eq:KP_Plucker}) 
are equivalent to (\ref{eq:KP_PluckerF}).
\end{demo}

\begin{corollary}
\label{cor:KP_PluckerF}
The function $\tau(x)$ is a solution of the KP hierarchy if and only if the coefficients $\xi_\lambda$ satisfy 
the following Pl\"ucker relations:
\begin{multline}
\label{eq:KP_PluckerF1}
\xi \begin{pmatrix} a_1, \dots, a_r \\ b_1, \dots, b_r \end{pmatrix}
\xi \begin{pmatrix} c_1, \dots, c_s \\ d_1, \dots, d_s \end{pmatrix}
\\
=
\sum_{k=1}^r
 (-1)^{r-k}
 \xi \begin{pmatrix} a_1, \dots, \widehat{a_k}, \dots, a_r \\ b_1, \dots, b_{r-1} \end{pmatrix}
 \xi \begin{pmatrix} a_k, c_1, \dots, c_s \\ b_r, d_1, \dots, d_s \end{pmatrix}
\\
+
\sum_{l=1}^s
 (-1)^{l-1}
 \xi \begin{pmatrix} a_1, \dots, a_r \\ b_1, \dots, b_{r-1}, d_l \end{pmatrix}
 \xi \begin{pmatrix} c_1, \dots, c_s \\ b_r, d_1, \dots, \widehat{d_l}, \dots, d_s \end{pmatrix},
\end{multline}
and
\begin{multline}
\label{eq:KP_PluckerF2}
\xi \begin{pmatrix} a_1, \dots, a_r \\ b_1, \dots, b_r \end{pmatrix}
\xi \begin{pmatrix} c_1, \dots, c_s \\ d_1, \dots, d_s \end{pmatrix}
\\
=
\sum_{k=1}^r
 (-1)^{r-k}
 \xi \begin{pmatrix} a_1, \dots, a_{r-1} \\ b_1, \dots, \widehat{b_k}, \dots, b_r \end{pmatrix}
 \xi \begin{pmatrix} a_r, c_1, \dots, c_s \\ b_k, d_1, \dots, d_s \end{pmatrix}
\\
+
\sum_{l=1}^s
 (-1)^{l-1}
 \xi \begin{pmatrix} a_1, \dots, a_{r-1}, c_l \\ b_1, \dots, b_r \end{pmatrix}
 \xi \begin{pmatrix} a_r, c_1, \dots, \widehat{c_l}, \dots, c_s \\ d_1, \dots, d_s \end{pmatrix},
\end{multline}
for any sequence of nonnegative integers $(a_1, \dots, a_r)$, $(b_1, \dots, b_r)$, $(c_1, \dots, c_s)$ 
and $(d_1, \dots, d_s)$.
\end{corollary}

\begin{remark}
We can show that the totality of the relations of the form (\ref{eq:KP_PluckerF1}) 
is equivalent to that of the relations of the form (\ref{eq:KP_PluckerF2}).
\end{remark}

\begin{demo}{Proof}
In order to derive (\ref{eq:KP_PluckerF1}), we consider the following substitution in (\ref{eq:KP_PluckerF}):
\begin{gather*}
(m_1, \dots, m_{p+1}) = (a_1, \dots, a_r),
\quad
(m'_1, \dots, m'_p) = (b_1, \dots, b_{r-1}),
\\
(n_1, \dots, n_q) = (c_1, \dots, c_s),
\quad
(n'_1, \dots, n'_{q+1}) = (b_r, d_1, \dots, d_s).
\end{gather*}
Then the left hand side of (\ref{eq:KP_PluckerF}) becomes
$$
\sum_{k=1}^r
 (-1)^k
 \xi \begin{pmatrix}
  a_1, \dots, \widehat{a_k}, \dots, a_r \\ b_1, \dots, b_{r-1}
 \end{pmatrix}
 \xi \begin{pmatrix}
  a_k, c_1, \dots, c_s \\ b_r, d_1, \dots, d_s
 \end{pmatrix},
$$
and the right hand side becomes
\begin{multline*}
(-1)^{(r-1)+1}
\xi \begin{pmatrix}
 a_1, \dots, a_r \\ b_1, \dots, b_{r-1}, b_r
\end{pmatrix}
\xi \begin{pmatrix}
 c_1, \dots, c_s \\ d_1, \dots, d_s
\end{pmatrix}
\\
+
\sum_{l=1}^s
 (-1)^{(r-1)+(l+1)}
 \xi \begin{pmatrix}
  a_1, \dots, a_r \\ b_1, \dots, b_{r-1}, d_l
 \end{pmatrix}
 \xi \begin{pmatrix}
  c_1, \dots, c_s \\ b_r, d_1, \dots, \widehat{d_l}, \dots, d_s
 \end{pmatrix}.
\end{multline*}
Hence we obtain (\ref{eq:KP_PluckerF1}).
Similarly, by considering the substitution
\begin{gather*}
(m_1, \dots, m_{p+1}) = (a_r, c_1, \dots, c_s),
\quad
(m'_1, \dots, m'_p) = (d_1, \dots, d_s),
\\
(n_1, \dots, n_q) = (a_1, \dots, a_{r-1}),
\quad
(n'_1, \dots, n'_{q+1}) = (b_1, \dots, b_r),
\end{gather*}
we obtain (\ref{eq:KP_PluckerF2}).
\end{demo}

\section{
Proof of Theorem~1.1
}

In this section we prove our main theorem (Theorem~\ref{thm:main}).

\subsection{%
Proof of Theorem~\ref{thm:main} (ii) $\implies$ (i)
}

First we prove a variant of the Pl\"ucker relation for determinants.
Given a matrix $Y$, a sequence of row indices $(i_1, \dots, i_k)$ 
and a sequence of column indices $(j_1, \dots, j_k)$, we put
$$
Y \begin{pmatrix} i_1, \dots, i_k \\ j_1, \dots, j_k \end{pmatrix}
 =
\left( y_{i_\alpha, j_\beta} \right)_{1 \le \alpha, \beta \le k}.
$$

\begin{prop}
\label{prop:det_PluckerF}
Let $Y$ be a matrix.
For row indices $m_1, \dots, m_{p+1}, n_1, \dots, n_q, l_1, \dots, l_r$ 
and column indices $m'_1, \dots, m'_p, n'_1, \dots, n'_{q+1}, l'_1, \dots, l'_r$, we have
\begin{multline}
\label{eq:det_PluckerF}
\sum_{\alpha=1}^{p+1}
 (-1)^\alpha
 \det Y \begin{pmatrix}
  m_1, \dots, \hat{m_\alpha}, \dots, m_{p+1}, l_1, \dots, l_r \\
  m'_1, \dots, m'_p, l'_1, \dots, l'_r
 \end{pmatrix}
 \det Y \begin{pmatrix}
  m_\alpha, n_1, \dots, n_q, l_1, \dots, l_r \\
  n'_1, \dots, n'_{q+1}, l'_1, \dots, l'_r
 \end{pmatrix}
\\
 =
\sum_{\beta=1}^{q+1}
 (-1)^{p+\beta}
 \det Y \begin{pmatrix}
  m_1, \dots, m_{p+1}, l_1, \dots, l_r \\
  m'_1, \dots, m'_p, n'_\beta, l'_1, \dots, l'_r
 \end{pmatrix}
 \det Y \begin{pmatrix}
  n_1, \dots, n_q, l_1, \dots, l_r \\
  n'_1, \dots, \hat{n'_\beta}, \dots, n'_{q+1}, l'_1, \dots, l'_r
 \end{pmatrix}.
\end{multline}
\end{prop}

\begin{demo}{Proof}
Recall the Pl\"ucker relation for Pfaffians (Ohta--Wenzel formula \cite{O,W}, see also \cite{IO, K}):
For a skew-symmetric matrix $X$ and row/column indices $i_1, \dots, i_p, j_1, \dots, j_q, k_1, \dots, k_r$, 
we have
\begin{multline}
\label{eq:Pf_Plucker}
\sum_{\alpha=1}^p
 (-1)^\alpha
 \Pf X(i_1, \dots, \hat{i_\alpha}, \dots, i_p, k_1, \dots, k_r)
 \Pf X(i_\alpha, j_1, \dots, j_q, k_1, \dots, k_r)
\\
=
\sum_{\beta=1}^q
 (-1)^\beta
 \Pf X(i_1, \dots, i_p, k_1, \dots, k_r, j_\beta)
 \Pf X(j_1, \dots, \hat{j_\beta}, \dots, j_q, k_1, \dots, k_r),
\end{multline}
where
$$
X(l_1, \dots, l_s)
 =
X \begin{pmatrix} l_1, \dots, l_s \\ l_1, \dots, l_s \end{pmatrix}.
$$
We apply this relation (\ref{eq:Pf_Plucker}) to the skew-symmetric matrix of the form
$$
X = \begin{pmatrix} O & Y \\ -\trans Y & O \end{pmatrix}
$$
with
\begin{gather*}
(i_1, \dots, i_p) \text{ replaced by } (m_1, \dots, m_{p+1}, m'_1, \dots, m'_p),
\\
(j_1, \dots, j_q) \text{ replaced by } (n_1, \dots, n_q, n'_1, \dots, n'_{q+1}),
\\
(k_1, \dots, k_r) \text{ replaced by } (l_1, \dots, l_r, l'_1, \dots, l'_r),
\end{gather*}
where $m_1, \dots, m_{p+1}, n_1, \dots, n_q, l_1, \dots, l_r$ are row indices of $Y$ in $X$ 
and $m'_1, \dots, m'_p, n'_1, \dots, n'_{q+1}, \allowbreak l'_1, \dots, l'_r$ are column indices of $Y$ in $X$.
Also we use the following relation between Pfaffians and determinants:
If $A$ is an $m \times n$ matrix, then we have
\begin{equation}
\label{eq:Pf=det}
\Pf \begin{pmatrix} O & A \\ -\trans A & O \end{pmatrix}
 =
\begin{cases}
 (-1)^{\binom{m}{2}} \det A &\text{if $m=n$,} \\
 0 &\text{if $m \neq n$.}
\end{cases}
\end{equation}

By using the alternating property of Pfaffians and (\ref{eq:Pf=det}), we have
\begin{align*}
&
\Pf X(m_1, \dots, \hat{m_\alpha}, \dots, m_{p+1}, m'_1, \dots, m'_p, l_1, \dots, l_r, l'_1, \dots, l'_r)
\\
&=
(-1)^{pr}
\Pf X(m_1, \dots, \hat{m_\alpha}, \dots, m_{p+1}, l_1, \dots, l_r, m'_1, \dots, m'_p, l'_1, \dots, l'_r)
\\
&=
(-1)^{pr} \cdot (-1)^{\binom{p+r}{2}}
\det Y \begin{pmatrix}
 m_1, \dots, \hat{m_\alpha}, \dots, m_{p+1}, l_1, \dots, l_r \\
 m'_1, \dots, m'_p, l'_1, \dots, l'_r
\end{pmatrix},
\end{align*}
and
\begin{align*}
&
\Pf X(m_\alpha, n_1, \dots, n_q, n'_1, \dots, n'_{q+1}, l_1, \dots, l_r, l'_1, \dots, l'_r)
\\
&=
(-1)^{(q+1)r}
\Pf X(m_\alpha, n_1, \dots, n_q, l_1, \dots, l_r, n'_1, \dots, n'_{q+1}, l'_1, \dots, l'_r)
\\
&=
(-1)^{(q+1)r} \cdot (-1)^{\binom{q+1+r}{2}}
\det Y \begin{pmatrix}
 m_\alpha, n_1, \dots, n_q, l_1, \dots, l_r \\
 n'_1, \dots, n'_{q+1}, l'_1, \dots, l'_r
\end{pmatrix}.
\end{align*}
Since $p+1+r>p-1+r$, we have
\begin{align*}
&
\Pf X(m_1, \dots, m_{p+1}, m'_1, \dots, \hat{m'_\alpha}, \dots, m'_p, l_1, \dots, l_r, l'_1, \dots, l'_r)
\\
&=
(-1)^{(p-1)r}
\Pf X(m_1, \dots, m_{p+1}, l_1, \dots, l_r, m'_1, \dots, \hat{m'_\alpha}, \dots, m'_p, l'_1, \dots, l'_r)
\\
&=0.
\end{align*}
Similarly we see that
\begin{align*}
&
\Pf X(n_1, \dots, \hat{n_\beta}, \dots, n_q, n'_1, \dots, n'_{q+1}, l_1, \dots, l_r, l'_1, \dots, l'_r)
\\
&\quad
=
0,
\\
&
\Pf X(m_1, \dots, m_{p+1}, m'_1, \dots, m'_p, l_1, \dots, l_r, l'_1, \dots, l'_r, n'_\beta)
\\
&\quad
=
(-1)^{pr+r} \cdot (-1)^{\binom{p+1+r}{2}}
\det Y \begin{pmatrix}
 m_1, \dots, m_{p+1}, l_1, \dots, l_r \\
 m'_1, \dots, m'_p, n'_\beta, l'_1, \dots, l'_r
\end{pmatrix},
\\
&
\Pf X(n_1, \dots, n_q, n'_1, \dots, \hat{n'_\beta}, \dots, n'_{q+1}, l_1, \dots, l_r, l'_1, \dots, l'_r)
\\
&\quad
=
(-1)^{qr} \cdot (-1)^{\binom{q+r}{2}}
\det Y \begin{pmatrix}
 n_1, \dots, n_q, l_1, \dots, l_r \\
 n'_1, \dots, \hat{n'_\beta}, \dots, n'_{q+1}, l'_1, \dots, l'_r
\end{pmatrix}.
\end{align*}
Hence, by canceling the common factor $(-1)^{pr + qr + r + \binom{p+r}{2} + \binom{q+r}{2} + q+r}$, 
we obtain the desired identity (\ref{eq:det_PluckerF}).
\end{demo}

\begin{corollary}
\label{cor:det_PluckerF}
Let $Y$ be a matrix.
For row indices $a_1, \dots, a_r, c_1, \dots, c_s, e_1, \dots, e_t$ 
and column indices $b_1, \dots, b_r, d_1, \dots, d_s, f_1, \dots, d_t$, we have
\begin{multline}
\label{eq:det_PluckerF1}
\det Y \begin{pmatrix} a_1, \dots, a_r, e_1, \dots, e_t \\ b_1, \dots, b_r, f_1, \dots, f_t \end{pmatrix}
\det Y \begin{pmatrix} c_1, \dots, c_s, e_1, \dots, e_t \\ d_1, \dots, d_s, f_1, \dots, f_t \end{pmatrix}
\\
=
\sum_{k=1}^r
 (-1)^{r-k}
 \det Y \begin{pmatrix} a_1, \dots, \hat{a_k}, \dots, a_r, e_1, \dots, e_t 
  \\ b_1, \dots, b_{r-1}, f_1, \dots, f_t \end{pmatrix}
 \det Y \begin{pmatrix} a_k, c_1, \dots, c_s, e_1, \dots, e_t 
  \\ b_r, d_1, \dots, d_s, f_1, \dots, f_t \end{pmatrix}
\\
+
\sum_{l=1}^s
 (-1)^{l-1}
 \det Y \begin{pmatrix} a_1, \dots, a_r, e_1, \dots, e_t 
  \\ b_1, \dots, b_{r-1}, d_l, f_1, \dots, f_t \end{pmatrix}
 \det Y \begin{pmatrix} c_1, \dots, c_s, e_1, \dots, e_t 
  \\ b_r, d_1, \dots, \hat{d_l}, \dots, d_s, f_1, \dots, f_t \end{pmatrix}.
\end{multline}
\end{corollary}

\begin{demo}{Proof}
We can use the same argument as in the derivation of Corollary~\ref{cor:KP_PluckerF} 
from Proposition~\ref{prop:KP_PluckerF}, so we omit the proof.
\end{demo}

Now we are in position to prove (ii) implies (i) in Theorem~\ref{thm:main}.

\begin{demo}{Proof of Theorem~\ref{thm:main} (ii) $\implies$ (i)}
Suppose that the coefficients $\xi_\lambda$ of $\tau(x)$ are expressed as determinants given in (\ref{eq:det}).
By applying the Pl\"ucker relation (\ref{eq:det_PluckerF1}) in Corollary~\ref{cor:det_PluckerF} 
to the matrices
$$
Y
 = 
\begin{pmatrix}
 \Big( z_{a,b} \Big)_{a, b \in \Int_{\ge 0}}
&
 \Big( u_a^{(j)} \Big)_{a \in \Int_{\ge 0}, 1 \le j \le s}
\\
 \Big( v_b^{(i)} \Big)_{1 \le i \le s, b \in \Int_{\ge 0}}
&
 O
\end{pmatrix},
$$
and its transpose, 
we see that $\xi_\lambda$ satisfy the Pl\"ucker relations (\ref{eq:KP_PluckerF1}) and (\ref{eq:KP_PluckerF2}) respectively,
hence $\tau(x) = \sum_\lambda \xi_\lambda s_\lambda(x)$ is a solution of the KP hierarchy.

Next we prove that $\xi_\lambda = 0$ unless $\lambda \supset \mu$.
We write $\lambda = (\alpha_1, \dots, \alpha_r | \beta_1, \dots, \beta_r)$ and 
$\mu = (\gamma_1, \dots, \gamma_s | \delta_1, \dots, \delta_s)$ in the Frobenius notation.
If $\lambda \not\supset \mu$, then at least one of the following holds:
\begin{enumerate}
\item[(a)]
$r < s$.
\item[(b)]
there exists an index $k$ such that $\alpha_k < \gamma_k$.
\item[(c)]
there exists an index $k$ such that $\beta_k < \delta_k$,
\end{enumerate}
If $r < s$, then $\xi_\lambda$ is the determinant of the $(r+s) \times (r+s)$ matrix, 
whose bottom-right block is the $s \times s$ zero matrix, so $\xi_\lambda = 0$.
If $\alpha_k < \gamma_k$, then we have
$$
\gamma_1 > \dots > \gamma_k > \alpha_k > \alpha_{k+1} > \dots > \alpha_r.
$$
Hence it follows from (\ref{eq:entry}) that $u_{\alpha_i}^{(j)} = 0$ for $k \le i \le r$ and $1 \le j \le k$, 
so $\xi_\lambda = 0$.
Similarly, if $\beta_k < \delta_k$, then we have $\xi_\lambda = 0$.

Finally we prove $\xi_\mu = 1$.
If $\lambda = \mu$, then we have
$$
\xi_\mu
 =
(-1)^s
\det
\begin{pmatrix}
 O
&
 \Big(
  \delta_{i,j} (-1)^{i-1}
 \Big)_{1 \le i, j \le s}
\\
 \Big(
  \delta_{i,j} (-1)^{i-1}
 \Big)_{1 \le i, j \le s}
&
 O
\end{pmatrix}
 =
1.
$$
Therefore $\tau(x)$ is a solution of the KP hierarchy satisfying (\ref{eq:cond1}) and (\ref{eq:cond2}).
\end{demo}

\begin{prop}
\label{prop:entry}
If the coefficients $\xi_\lambda$ are defined by (\ref{eq:det}), then the entries of the matrix in (\ref{eq:det}) 
are given by some specific coefficients as follows:
\begin{align*}
z_{a,b}
 &=
\xi \begin{pmatrix} a, \gamma_1, \dots, \gamma_s \\ b, \delta_1, \dots, \delta_s \end{pmatrix},
\\
u_a^{(j)}
 &=
\xi \begin{pmatrix} a, \gamma_1, \dots, \hat{\gamma_j}, \dots, \gamma_s \\ \delta_1, \dots, \delta_s \end{pmatrix},
\\
v_b^{(i)}
 &=
\xi \begin{pmatrix} \gamma_1, \dots, \gamma_s \\ b, \delta_1, \dots, \hat{\delta_i}, \dots, \delta_s \end{pmatrix}.
\end{align*}
\end{prop}

\begin{demo}{Proof}
By the skew-symmetry of $\xi \begin{pmatrix} \alpha_1, \dots, \alpha_r \\ \beta_1, \dots, \beta_r \end{pmatrix}$ 
and determinants, we have
$$
\xi \begin{pmatrix} \alpha_1, \dots, \alpha_r \\ \beta_1, \dots, \beta_r \end{pmatrix}
 =
(-1)^s
\det
\begin{pmatrix}
 \big( z_{\alpha_i, \beta_j} \big)_{1 \le i, j \le r}
&
 \big( u_{\alpha_i}^{(j)} \big)_{1 \le i \le r, 1 \le j \le s}
\\
 \big( v_{\beta_j}^{(i)} \big)_{1 \le i \le s, 1 \le j \le r}
&
 O
\end{pmatrix},
$$
Hence, by using the conditions given in (\ref{eq:entry}), we see that
\begin{align*}
&
\xi \begin{pmatrix} a, \gamma_1, \dots, \gamma_s \\ b, \delta_1, \dots, \delta_s \end{pmatrix}
\\
 &=
(-1)^s \det \left( \begin{array}{cccc|ccc}
 z_{a,b} & 0 & \dots & 0  & u_a^{(1)} & \dots & u_a^{(s)} \\
 0       &   &        &    & 1         &        &           \\
 \vdots  &   & O      &    &           & \ddots &           \\
 0       &   &        &    &           &        & (-1)^{s-1} \\
\hline
 v_b^{(1)} & 1 &      &    &           &        & \\
 \vdots    &   & \ddots &  &           & O      & \\
 v_b^{(s)} &   &        & (-1)^{s-1} & &        &
\end{array} \right)
\\
&=
z_{a,b},
\end{align*}
and
\begin{align*}
&
\xi \begin{pmatrix} a, \gamma_1, \dots, \hat{\gamma_j}, \dots, \gamma_s \\ \delta_1, \dots, \delta_s \end{pmatrix}
\\
&=
(-1)^s \det \left( \begin{array}{ccc|ccccccc}
   &        &            & u_a^{(1)} & \dots & u_a^{(j-1)} & u_a^{(j)} & u_a^{(j+1)} & \dots & u_a^{(s)} \\
   &        &            & 1         &        &             & 0         &             &        &           \\
   &        &            &           & \ddots &             & \vdots    &             &        &           \\
   & O      &            &           &        & (-1)^{j-2}  & 0         &             &        &           \\
   &        &            &           &        &             & 0         & (-1)^j      &        &           \\
   &        &            &           &        &             & \vdots    &             & \ddots &           \\
   &        &            &           &        &             & 0         &             &        & (-1)^{s-1} \\
\hline
 1 &        &            & \\
   & \ddots &            &           &        &             & O         & \\
   &        & (-1)^{s-1} & &        &
\end{array} \right)
\\
 &=
u_a^{(j)}.
\end{align*}
\end{demo}

\subsection{%
Proof of Theorem~\ref{thm:main} (i) $\implies$ (ii)
}

First we show that a solution of the KP hierarchy are determined by a part of its coefficients.

\begin{lemma}
\label{lem:polynomial}
Fix a partition $\mu$.
Suppose that $\tau(x) = \sum_\lambda \xi_\lambda s_\lambda(x)$ is a solution of the KP hierarchy 
and satisfies the conditions (\ref{eq:cond1}) and (\ref{eq:cond2}).
Then $\xi_\lambda$ can be expressed as a polynomial in 
\begin{align*}
I_\mu
 &=
\left\{
 \xi \begin{pmatrix} a, \gamma_1, \dots, \gamma_s \\ b, \delta_1, \dots, \delta_s \end{pmatrix}
 : a, b \in \Int_{\ge 0}
\right\}
\\
 &\quad
 \cup
\left\{
 \xi \begin{pmatrix} a, \gamma_1, \dots, \hat{\gamma_j}, \dots, \gamma_s \\ \delta_1, \dots, \delta_s \end{pmatrix}
 : a \in \Int_{\ge 0}, 1 \le j \le s
\right\}
\\
 &\quad
 \cup
\left\{
 \xi \begin{pmatrix} \gamma_1, \dots, \gamma_s \\ b, \delta_1, \dots, \hat{\delta_i}, \dots, \delta_s \end{pmatrix}
 : b \in \Int_{ge 0}, 1 \le i \le s
\right\}.
\end{align*}
\end{lemma}

\begin{demo}{Proof}
Proceed by the induction on $p(\lambda) = \# \{ i : \lambda_i \ge i \}$ and $|\lambda|$.
If $\lambda = \emptyset$, then we have
$$
\xi_\lambda
 = 
\begin{cases}
 1 &\text{if $\mu = \emptyset$,} \\
 0 &\text{if $\mu \neq \emptyset$.}
\end{cases}
$$

Let $\lambda = (\alpha_1, \dots, \alpha_r|\beta_1, \dots, \beta_r)$ be a partition.
If $\{ \alpha_1, \dots, \alpha_r \} \subset \{ \gamma_1, \dots, \gamma_s \}$ 
and $\{ \beta_1, \dots, \beta_r \} \subset \{ \delta_1, \dots, \delta_s \}$, 
then $|\lambda| \le |\mu|$ and 
$$
\xi_\lambda
 = 
\begin{cases}
 1 &\text{if $\lambda = \mu$,} \\
 0 &\text{if $\lambda \neq \mu$.}
\end{cases}
$$
Next we consider the case where $\{ \beta_1, \dots, \beta_r \} \not\subset \{ \delta_1, \dots, \delta_s \}$.
In this case, if we put
$$
(\beta'_1, \dots, \beta'_r) = (\beta_1, \dots, \hat{\beta_i}, \dots, \beta_r, \beta_i),
$$
then we have
$$
\xi_\lambda
 =
(-1)^{r-i}
\xi \begin{pmatrix} \alpha_1, \dots, \alpha_r \\ \beta'_1, \dots, \beta'_r \end{pmatrix}.
$$
Then, by using the assumption (\ref{eq:cond2}) and the Pl\"ucker relation (\ref{eq:KP_PluckerF1}), we have
\begin{align*}
\xi_\lambda
 &=
(-1)^{r-i} 
\xi \begin{pmatrix} \alpha_1, \dots, \alpha_r \\ \beta'_1, \dots, \beta'_r \end{pmatrix}
\xi \begin{pmatrix} \gamma_1, \dots, \gamma_s \\ \delta_1, \dots, \delta_s \end{pmatrix}
\\
 &=
(-1)^{r-i}
\left\{
\begin{array}{l}
 \displaystyle\sum_{k=1}^r
  (-1)^{r-k}
  \xi \begin{pmatrix} \alpha_1, \dots, \hat{\alpha_k}, \dots, \alpha_r \\ \beta'_1, \dots, \beta'_{r-1} \end{pmatrix}
  \xi \begin{pmatrix} \alpha_k, \gamma_1, \dots, \gamma_s \\ \beta'_r, \delta_1, \dots, \delta_s \end{pmatrix}
\\
+
(-1)^{r-i}
 \displaystyle\sum_{l=1}^s
  (-1)^{l-1}
  \xi \begin{pmatrix} \alpha_1, \dots, \alpha_r \\ \beta'_1, \dots, \beta'_{r-1}, \delta_l \end{pmatrix}
  \xi \begin{pmatrix} \gamma_1, \dots, \gamma_s \\ \beta'_r, \delta_1, \dots, \hat{\delta_l}, \dots, \delta_s \end{pmatrix}
\end{array}
\right\}.
\end{align*}
By the induction hypothesis on $p(\lambda) = r$, we see that 
$\xi \begin{pmatrix} \alpha_1, \dots, \hat{\alpha_k}, \dots, \alpha_r \\ \beta'_1, \dots, \beta'_{r-1} \end{pmatrix}$ 
is a polynomial in $I_\mu$.
If $\delta_l < \beta'_r = \beta_i$, 
then $|\alpha| + |\beta| - \beta'_r +\delta_l < |\alpha| + |\beta|$ and 
it follows from the induction hypothesis on $|\lambda|$ that 
$\xi \begin{pmatrix} \alpha_1, \dots, \alpha_r \\ \beta'_1, \dots, \beta'_{r-1}, \delta_l \end{pmatrix}$ 
is a polynomial in $I_\mu$.
If $\delta_l > \beta'_r = \beta_i$, 
then $|\gamma|+|\delta| - \delta_l + \beta'_r < |\gamma| + |\delta|$ and 
it follows from the assumption (\ref{eq:cond1}) that
 $\xi \begin{pmatrix} \gamma_1, \dots, \gamma_s \\ \beta'_r, \delta_1, \dots, \hat{\delta_j}, \dots, \delta_s 
\end{pmatrix} = 0$.
Therefore $\xi_\lambda$ is a polynomial in $I_\mu$.

If $\{ \alpha_1, \dots, \alpha_r \} \not\subset \{ \gamma_1, \dots, \gamma_s \}$, then 
we can use the Pl\"ucker relation (\ref{eq:KP_PluckerF2}) to prove that $\xi_\lambda$ is a polynomial in $I_\mu$.
\end{demo}

\begin{demo}{Proof of Theorem~\ref{thm:main} (i) $\implies$ (ii)}
Suppose that $\tau(x)$ is a solution of the KP hierarchy satisfying the conditions (\ref{eq:cond1}) and (\ref{eq:cond2}).
For a partition $\lambda$, we define $\xi'_\lambda$ by putting
\begin{multline*}
\xi'_\lambda
\\
 =
(-1)^s
\det
\begin{pmatrix}
 \left(
  \xi \begin{pmatrix} \alpha_i,\gamma_1, \dots, \gamma_s \\ \beta_j, \delta_1, \dots, \delta_s \end{pmatrix}
 \right)_{1 \le i, j \le r}
&
 \left(
  \xi \begin{pmatrix} \alpha_i, \gamma_1, \dots, \hat{\gamma_j}, \dots, \gamma_s \\ \delta_1, \dots, \delta_s \end{pmatrix}
 \right)_{1 \le i \le r, 1 \le j \le s}
\\
 \left(
  \xi \begin{pmatrix} \gamma_1, \dots, \gamma_s \\ \beta_j, \delta_1, \dots, \hat{\delta_i}, \dots, \delta_s \end{pmatrix}
 \right)_{1 \le i \le s, 1 \le j \le r}
&
 O
\end{pmatrix}.
\end{multline*}
Then we have shown in the proof of (ii) $\implies$ (i) 
that $\sum_\lambda \xi'_\lambda s_\lambda(x)$ is a solution of the KP hierarchy satisfying the conditions 
(\ref{eq:cond1}) and (\ref{eq:cond2}).
Also by Proposition~\ref{prop:entry} we see that
\begin{gather*}
\xi \begin{pmatrix} a, \gamma_1, \dots, \gamma_s \\ b, \delta_1, \dots, \delta_s \end{pmatrix}
 =
\xi' \begin{pmatrix} a, \gamma_1, \dots, \gamma_s \\ b, \delta_1, \dots, \delta_s \end{pmatrix},
\\
\xi \begin{pmatrix} a, \gamma_1, \dots, \hat{\gamma_j}, \dots, \gamma_s \\ \delta_1, \dots, \delta_s \end{pmatrix}
 =
\xi' \begin{pmatrix} a, \gamma_1, \dots, \hat{\gamma_j}, \dots, \gamma_s \\ \delta_1, \dots, \delta_s \end{pmatrix},
\\
\xi \begin{pmatrix} \gamma_1, \dots, \gamma_s \\ b, \delta_1, \dots, \hat{\delta_i}, \dots, \delta_s \end{pmatrix}
 =
\xi' \begin{pmatrix} \gamma_1, \dots, \gamma_s \\ b, \delta_1, \dots, \hat{\delta_i}, \dots, \delta_s \end{pmatrix}.
\end{gather*}
Now it follows from Lemma~\ref{lem:polynomial} that $\xi_\lambda = \xi'_\lambda$.
This completes the proof of Theorem~\ref{thm:main}.
\end{demo}

\section{
Relation with the Sato Grassmannian
}

By Sato's theory on the KP hierarchy, the expansion coefficients of the $\tau$-function are given 
by Pl\"ucker coordinates of a point of the Sato Grassmannian $\UGM$.
In this section we show that the formulas for Pl\"ucker coordinates coincide with those given 
in (\ref{eq:det}) of Theorem~\ref{thm:main} (ii).

Let $V = \Comp((z))$ be the vector space of formal Laurent series in the variable $z$ 
and $V_\emptyset = \Comp[z^{-1}]$, $V_0 = z \Comp[[z]]$ subspaces of $V$.
Then 
$$
V = V_\emptyset \oplus V_0.
$$
Let $\pi: V \rightarrow V/V_0 \simeq V_\emptyset$ be the natural projection. 
Then the Sato Grassmannian $\UGM$ is defined as the set of subspaces $U$ of $V$ 
such that $\dim {\mathrm Ker}(\pi|_U) = \dim {\mathrm Coker}(\pi|_U) < \infty$.
The Sato Grassmannian has the cell decomposition 
$$
\UGM = \bigsqcup_\mu \UGM^\mu,
$$
where $\mu$ runs over all partitions.
The cell $\UGM^\mu$ associated with a partition $\mu$ is described in the following manner.

We write an element $f$ of $V$ as
$$
f = \sum_{a \in \Int} X_a z^{a+1}.
$$
and identify $f$ with the infinite column vector $(X_a)_{a \in \Int}$.
A point $U$ of $\UGM$ is represented by its ordered basis, called a frame, 
which is viewed as a matrix $\Xi = \left( X_{a,j} \right)_{a \in \Int, j \ge 1}$ 
with $\left( X_{a,j} \right)_{a \in \Int}$ the $j$th basis vector.
To each $U \in \UGM$, we can associate the Maya diagram $M$ of charge $0$ 
and the frame $\Xi$ uniquely determined by the following condition
\begin{equation}
\label{eq:frame}
X_{a,j}
 =
\begin{cases}
1 & \text{if $a = m_j$,} \\
0 & \text{if $a < m_j$ or $a = m_k$ for some $k < j$,}
\end{cases}
\end{equation}
where $m_i$ is the $i$th largest element of $M$.
We call such a frame $\Xi=\Xi(U)$ the normalized frame of $U$. 
The cell $\UGM^\mu$ is the set of points of $\UGM$ corresponding to the Maya diagram $M$ 
associated with $\mu$.

Let $U \in \UGM^\mu$ and $\Xi$ the normalized frame of $U$.
Given a partition $\lambda$,
the Pl\"ucker coordinate $\xi^U_\lambda$ of $\Xi$ is defined as 
\begin{equation}
\label{eq:Pcoord}
\xi^U_\lambda
 =
\det \left( X_{l_i,j} \right)_{i, j \ge 1},
\end{equation}
where $L = (l_1, l_2, \dots)$ is the Maya diagram corresponding to $\lambda$.
Although this is an infinite determinant it actually is defined as a finite determinant as follows.
Take any positive integer $N$ such that $l_i = -i$ holds for any $i \ge N$.
Then 
$$
\xi^U_\lambda
 =
\det \left( X_{l_i,j} \right)_{1 \le i, j \le N},
$$
which does not depend on the choice of $N$. 

For a point $U \in \UGM$, we define the corresponding $\tau$-function $\tau^U(x)$ by
\begin{equation}
\label{eq:tau-U}
\tau^U(x)
 =
\sum_\lambda \xi^U_\lambda s_\lambda(x).
\end{equation}
Then 

\begin{theorem}
\label{thm:Sato1}
(\cite{SS})
For a point $U \in \UGM$, 
the $\tau$-function $\tau^U(x)$ is a solution of the KP hierarchy. 
Conversely for any formal power series solution $\tau(x)$ of the KP hierarchy,
there is a point $U$ of $\UGM$ and a constant $C$ such that 
$\tau(x) = C \tau^U(x)$.
\end{theorem}

\begin{theorem}
\label{thm:Sato2}
(\cite{SS})
The set of Pl\"ucker coordinates $\{ \xi^U_\lambda \}_\lambda$ of a point of $\UGM$ 
satisfies the Pl\"ucker relations (\ref{eq:KP_Plucker}) in Proposition~\ref{prop:KP_Plucker}.
Conversely, for any $\{ \xi_\lambda \}_\lambda$ satisfying the Pl\"ucker relations (\ref{eq:KP_Plucker})
there exists a point $U \in \UGM$ such that $\xi_\lambda$ is the Pl\"ucker coordinate of 
$U$ for any partition $\lambda$.
\end{theorem}

Proposition~\ref{prop:KP_Plucker} is a consequence of these theorems.

Now we consider the Pl\"ucker coordinates of a point $U$ in the cell $\UGM^\mu$ associated to a partition $\mu$.

\begin{lemma}
\label{lem:Pcoord_cond}
Let $\mu$ be a partition and $U \in \UGM^\mu$.
Then the Pl\"ucker coordinates $\{ \xi^U_\lambda \}_\lambda$ satisfy
$\xi^U_\mu = 1$ and $\xi^U_\lambda = 0$ unless $\lambda \supset \mu$.
\end{lemma}

\begin{demo}{Proof}
Let $M = (m_1, m_2, \dots)$ be the Maya diagram corresponding to $\mu$ 
and $\Xi = (X_{a,j})$ the normalized frame of $U$.
Since the matrix $\left( X_{m_i}, j \right)_{1 \le i, j \le N}$ is the identity matrix, 
we have $\xi^U_\mu = 1$.
If $\lambda \not\supset \mu$, then there is an index $k$ such that $\lambda_k < \mu_k$,
and we see that $X_{l_i,j} = 0$ for $i \ge k$ and $j \le k$, which implies $\xi^U_\lambda = 0$.
\end{demo}

The entries of the normalized frame of a point of $\UGM^\mu$ are expressed as special Pl\"ucker coordinates.

\begin{prop}
\label{prop:frame_entry}
Let $\mu = (\gamma_1, \dots, \gamma_s | \delta_1, \dots, \delta_s)$ be a partition and $U \in \UGM^\mu$.
Then the entries $X_{a,j}$ ($a \in \Int$, $j \ge 1$) of the normalized frame of $U$ 
are given as follows:
\begin{enumerate}
\item[(i)]
If $m_j \ge 0$, then $m_j = \gamma_j$ and we have
\begin{equation}
\label{eq:frame_entry1}
X_{a,j}
 =
\begin{cases}
0 &\text{if $a < \gamma_j$,} \\
(-1)^{j-1} &\text{if $a = \gamma_j$,} \\
(-1)^{j-1}
 \xi^U
  \begin{pmatrix}
   a, \gamma_1, \dots, \hat{\gamma_j}, \dots, \gamma_s \\
   \delta_1, \dots, \delta_s
  \end{pmatrix}
 &\text{if $a > \gamma_j$.}
\end{cases}
\end{equation}
\item[(ii)]
If $m_j<0$ and $m_j = -b-1$ with $b \ge 0$, then we have
\begin{equation}
\label{eq:frame_entry2}
X_{a,j}
=
\begin{cases}
0 &\text{if $a < m_j$, or if $a > m_j$ and $a \in M$,} \\
1 &\text{if $a = m_j$,} \\
(-1)^{b-\delta_i-i}
 \xi^U
  \begin{pmatrix}
   \gamma_1, \dots, \gamma_s \\
   b, \delta_1, \dots, \hat{\delta_i}, \dots, \delta_s
  \end{pmatrix}
 &\text{if $a > m_j$ and $a = -\delta_i-1$ for some $i$,} \\
(-1)^b
 \xi^U
  \begin{pmatrix}
   a, \gamma_1, \dots, \gamma_s\\
   b, \delta_1, \dots, \delta_s
  \end{pmatrix}
 &\text{if $a \ge 0$.}
\end{cases}
\end{equation}
\end{enumerate}
\end{prop}

\begin{demo}{Proof}
The Pl\"ucker coordinate $\xi^U_\lambda$ of $U$ can be computed 
by using the semi-infinite wedge product in the following way.
Let $\be_j = ( \delta_{ij} )_{i \ge 1}$ be the infinite unit row vector with $1$ at the $j$th position. 
Given the normalized frame $\Xi = (X_{a,j})$ of a point $U \in \UGM^\mu$, 
we put
$$
\bv_a = \sum_{j=1}^\infty X_{a,j} \be_j \quad(a \in \Int).
$$
In particular $\bv_{m_k} = \be_k$ for any $k \ge 1$.
If a partition $\lambda$ corresponds to the Maya digram $L = ( l_1, l_2, \dots )$, then we have
$$
\bv_{l_1} \wedge \bv_{l_2} \wedge \dots
 =
\xi^U_\lambda \be_1 \wedge \be_ 2\wedge \cdots.
$$

(i)
The first two case $a \le \gamma_j$ are obvious from the definition (\ref{eq:frame}) of $X_{a,j}$.
So we assume $a > \gamma_j$.
Let $k$ be the largest index such that $a < \gamma_k$ (if $a > \gamma_1$ set $k=0$). 
Since $a > \gamma_j$, we have $k < j$.
Define a partition $\lambda$ by
$$
\lambda
 =
(\gamma_1, \dots, \gamma_k, a, \gamma_{k+1}, \dots, \hat{\gamma_j}, \dots, \gamma_s
 | \delta_1, \dots, \delta_s).
$$
Then the corresponding Maya diagram $L$ is given as
$$
L
 = (m_1, \dots, m_k, a, m_{k+1}, \dots, m_{j-1}, m_{j+1}, \dots).
$$
Since
\begin{align*}
\bv_{l_1} \wedge \bv_{l_2} \wedge \cdots
&=
\be_1 \wedge \cdots \wedge \be_k \wedge \bv_a \wedge \be_{k+1} \wedge \cdots
 \wedge \be_{j-1} \wedge \be_{j+1} \wedge \cdots
\\
&=
X_{a,j}
\be_1 \wedge \cdots \wedge \be_k \wedge \be_j \wedge \be_{k+1} \wedge \cdots
 \wedge \be_{j-1} \wedge \be_{j+1} \wedge \cdots
\\
&=
(-1)^{j-1-k} X_{a,j}
\be_1\wedge \be_2 \wedge \cdots,
\end{align*}
we have
$$
\xi^U_\lambda = (-1)^{j-1-k} X_{a,j}.
$$
On the other hand, we have
$$
\xi^U_\lambda
 =
\xi^U
 \begin{pmatrix}
  \gamma_1, \dots, \gamma_k, a, \gamma_{k+1}, \dots, \hat{\gamma_j}, \dots, \gamma_s \\
  \delta_1, \dots, \delta_s
 \end{pmatrix}
=
(-1)^k
\xi^U
 \begin{pmatrix}
  a, \gamma_1, \dots, \hat{\gamma_j}, \dots, \gamma_s \\
  \delta_1, \dots, \delta_s
 \end{pmatrix}.
$$
Hence we have
$$
X_{a,j}
 = 
(-1)^{j-1}
\xi^U \begin{pmatrix}
  a, \gamma_1, \dots, \hat{\gamma_j}, \dots, \gamma_s \\
  \delta_1, \dots, \delta_s
 \end{pmatrix}.
$$

(ii)
The first two cases follow from the definition (\ref{eq:frame}) of the normalized frame. 

Let us prove the third case.
Assume $a > m_j$ and $a = -\delta_i-1$ for some $i$.
Since $a \not\in M$, there exists an index $k$ such that $m_k > a > m_{k+1}$. 
Let $l$ be the largest index such $b < \delta_l$ (if $b > \delta_1$ set $l=0$).
Since $a > m_j$, we have $k < j$ and $l<i$.
Define a partition $\lambda$ by
$$
\lambda
 =
(\gamma_1, \dots, \gamma_s 
| \delta_1, \dots, \delta_l, b, \delta_{l+1}, \dots, \hat{\delta_i}, \dots, \delta_s).
$$
Then the corresponding Maya diagram $L = (l_1, l_2, \dots)$ is given by
$$
L = (m_1, \dots, m_k, a, m_{k+1}, \dots, m_{j-1}, m_{j+1}, \dots).
$$
By the same argument as in the proof of (i), we have
$$
\xi_\lambda^U
 =
(-1)^{j-1-k} X_{a,j}.
$$
Here, by using $m_j = -b-1$, $a = -\delta_i-1$ and the definition of the index $l$, we see that
$$
j-1-k
 =
\# \{ m \in M : m_j < m < a \}
 =
(-\delta_i-2) - (-b-1) - (i-1-l)
 =
-\delta_i + b - i + l.
$$
On the other hand 
$$
\xi^U_\lambda
 =
\xi^U
 \begin{pmatrix}
  \gamma_1, \dots, \gamma_s \\
  \delta_1, \dots, \delta_l, b, \delta_{l+1}, \dots, \hat{\delta_i}, \dots, \delta_s
 \end{pmatrix}
 =
(-1)^l
 \xi^U
  \begin{pmatrix}
   \gamma_1, \dots, \gamma_s \\
   b, \delta_1, \dots, \hat{\delta_i}, \dots, \delta_s
  \end{pmatrix}.
$$
Hence we have
$$
X_{a,j}
 =
(-1)^{b-\delta_i-i}
\xi^U
 \begin{pmatrix}
  \gamma_1, \dots, \gamma_s \\
  b, \delta_1, \dots, \hat{\delta_i}, \dots, \delta_s
 \end{pmatrix}.
$$

Finally we prove the fourth case.
Suppose that $a \ge 0$.
Let $k$ be the largest index such that $a < \gamma_k$ (if $a > \gamma_1$ set $k=0$), 
and let $l$ be the largest index such that $b < \delta_l$ (if $b > \delta_1$ set $l=0$).
Define a partition $\lambda$ by
$$
\lambda
 =
(\gamma_1, \dots, \gamma_k, a, \gamma_{k+1}, \dots, \gamma_s
 | \delta_1, \dots, \delta_l, b, \delta_{l+1}, \dots, \delta_s)
$$
then the corresponding Maya diagram $L = (l_1, l_2, \dots)$ is given by
$$
L
 = (m_1, \dots, m_k, a, m_{k+1}, \dots, m_{j-1}, m_{j+1}, \dots).
$$
By the same argument as in the proof of (i), we have
$$
\xi_\lambda^U
 =
(-1)^{j-1-k} X_{a,j}.
$$
Here, by using $m_j = -b-1$ and the definition of the indices $k$ and $l$, we see that 
$$
j-1-k
 =
\# \{ m \in M : m_j < m < a \}
 =
s-k+b-(s-l)
 =
b-k+l.
$$
On the other hand, we have 
$$
\xi^U_\lambda
 =
\xi^U
 \begin{pmatrix}
  \gamma_1, \dots, \gamma_k, a, \gamma_{k+1}, \dots, \gamma_s \\
  \delta_1, \dots, \delta_l, b, \delta_{l+1}, \dots, \delta_s
 \end{pmatrix}
 =
(-1)^{k+l}
\xi^U
 \begin{pmatrix}
  a, \gamma_1, \dots, \gamma_s \\
  b, \delta_1, \dots, \delta_s
 \end{pmatrix}.
$$
Hence we have
$$
X_{a,j}
 =
(-1)^b
\xi^U
 \begin{pmatrix}
  a, \gamma_1, \dots, \gamma_s \\
  b, \delta_1, \dots, \delta_s
 \end{pmatrix}.
$$
\end{demo}

Notice that the Pl\"ucker coordinates appearing in the entries of the normalized frame 
given in Proposition~\ref{prop:frame_entry} are exactly the same as $I_\mu$ in Lemma~\ref{lem:polynomial} 
(except $0$ and $\pm 1$ forced by the conditions given in (\ref{eq:entry})).
In other words the variables in $I_\mu \backslash \{ 0, \pm 1 \}$ form the affine coordinates of $\UGM^\mu$.
So an arbitrary Pl\"ucker coordinate can be expressed by elements in $I_\mu$.

If $U$ is a point in the cell $\UGM^\mu$ associated to a partition $\mu$, 
then by Theorem~\ref{thm:Sato1} and Lemma~\ref{lem:Pcoord_cond} 
the corresponding $\tau$-function $\tau^U(x)$ (\ref{eq:tau-U}) 
is a solution of the KP hierarchy satisfying the conditions (\ref{eq:cond1}) and (\ref{eq:cond2}) 
in Theorem~\ref{thm:main}.
Thus the following proposition is a corollary of Theorem~\ref{thm:main} and Proposition~\ref{prop:entry}.

\begin{prop}
\label{prop:Pcoord_det}
Let $\mu = (\gamma_1, \dots, \gamma_s | \delta_1, \dots, \delta_s)$ be a partition 
and $U \in \UGM^\mu$.
Then the Pl\"ucker coordinate $\xi^U_\lambda$ corresponding to a partition 
$\lambda = (\alpha_1, \dots, \alpha_r | \beta_1, \dots, \beta_r)$ is expressed as
\begin{equation}
\label{eq:Pcoord_det}
\xi^U_{\lambda}
 =
(-1)^s
\det \begin{pmatrix}
\left( 
 \tilde{z}_{\alpha_i,\beta_j}
\right)_{1 \le i, j \le r}
&
\left( 
 \tilde{u}^{(j)}_{\alpha_i}
\right)_{1 \le i \le r, 1 \le j \le s}
\\
\left( 
 \tilde{v}^{(i)}_{\beta_j}
\right)_{1 \le i \le s, 1 \le j \le r}
&
 O
\end{pmatrix},
\end{equation}
where
\begin{align*}
\tilde{z}_{a,b}
 &=
\xi^U
 \begin{pmatrix}
  a, \gamma_1, \dots, \gamma_s \\
  b, \delta_1, \dots, \delta_s
 \end{pmatrix},
\\
\tilde{u}^{(j)}_a
 &=
\xi^U
 \begin{pmatrix}
  a, \gamma_1, \dots, \hat{\gamma}_j, \dots, \gamma_s \\
  \delta_1, \dots, \delta_s
 \end{pmatrix},
\\
\tilde{v}^{(i)}_b
 &=
\xi^U
 \begin{pmatrix}
  \gamma_1, \dots, \gamma_s \\
  b, \delta_1, \dots, \hat{\delta}_i, \dots, \delta_s
 \end{pmatrix}.
\end{align*}
\end{prop}

We can compute the Pl\"ucker coordinates of a point of $\UGM^\mu$ 
by using the definition (\ref{eq:Pcoord}) and Proposition~\ref{prop:frame_entry} 
and show that the formula (\ref{eq:Pcoord_det}) is naturally obtained.

Here we assume that two partitions $\lambda = (\alpha_1, \dots, \alpha_r|\beta_1, \dots, \beta_r)$ 
and $\mu = (\gamma_1, \dots, \gamma_s|\delta_1, \dots, \delta_s)$ satisfy the condition
\begin{equation}
\label{eq:cond_generic}
\{ \beta_1, \dots, \beta_r \} \cap \{ \delta_1, \dots, \delta_s \} = \emptyset
\end{equation}
and compute the Pl\"ucker coordinate $\xi^U_\lambda$ of $U \in \UGM^\mu$.
Put
$$
n_i = \# \{ j : \beta_i > \delta_j \},
\quad
n'_i = \# \{ j : \beta_j < \delta_i \}.
$$
Let $M = (m_1, m_2, \dots)$ and $L = (l_1, l_2, \dots)$ be the Maya diagrams corresponding to $\mu$ and $\lambda$
respectively.
We have
$$
L \cap \Int_{<0}
 =
\left( ( M \cap \Int_{<0} ) \setminus \{ - \beta_1-1, \dots, - \beta_r-1 \} \right) 
 \sqcup \{ -\delta_1-1, \dots, -\delta_s-1 \}.
$$
Let $k_j$ be the index such that $m_{k_j} = - \beta_j - 1$.
Then we have
\begin{multline*}
\bv_{l_1} \wedge \bv_{l_2} \wedge \cdots
\\
 =
(-1)^A 
 \bv_{\alpha_1} \wedge \cdots \wedge \bv_{\alpha_r} \wedge
 \bv_{-\delta_s-1} \wedge \cdots \wedge \bv_{-\delta_1-1}
 \wedge \be_{s+1} \wedge \be_{s+2} \wedge \cdots \wedge \hat{\be_{k_r}} \wedge \cdots
 \wedge \hat{\be_{k_1}} \wedge \cdots
\end{multline*}
with
$$
A = \sum_{j=1}^s (\delta_j-n'_j-(s-j)) = \sum_{j=1}^s \delta_j - \sum_{j=1}^s n'_j - \binom{s}{2}.
$$
Since we have
\begin{multline*}
\bv_{\alpha_1} \wedge \cdots \wedge \bv_{\alpha_r} \wedge
 \bv_{-\delta_s-1} \wedge \cdots \wedge \bv_{-\delta_1-1}
\\
 =
\sum_{c_i,d_j}
 X_{\alpha_1,c_1} \cdots X_{\alpha_r,c_r} X_{-\delta_s-1,d_s} \cdots X_{-\delta_1-1,d_1}
\be_{c_1} \wedge \cdots \be_{c_r} \wedge\be_{d_s} \wedge \cdots\be_{d_1},
\end{multline*}
and
$$
\be_{c_1} \wedge \cdots \be_{c_r} \wedge\be_{d_s} \wedge \cdots\be_{d_1} \wedge
\be_{s+1} \wedge \be_{s+2} \wedge \cdots \wedge \hat{\be_{k_r}} \wedge \cdots
\wedge \hat{\be_{k_1}} \wedge \cdots
 =
0
$$
unless $\{ c_1, \dots, c_r, d_s, \dots, d_1 \} = \{ 1, \dots, s, k_r, \dots, k_1 \}$, 
we see that
\begin{align*}
&
\bv_{\alpha_1} \wedge \cdots \wedge \bv_{\alpha_r} \wedge
 \bv_{-\delta_s-1} \wedge \cdots \wedge \bv_{-\delta_1-1} \wedge
\be_{s+1} \wedge \be_{s+2} \wedge \cdots \wedge \hat{\be_{k_r}} \wedge \cdots
\wedge \hat{\be_{k_1}} \wedge \cdots
\\
&\quad
 =
\det (G)
\be_{1} \wedge \cdots \wedge \be_{s} \wedge \be_{k_r} \wedge \cdots \wedge \be_{k_1} \wedge
\be_{s+1} \wedge \be_{s+2} \wedge \cdots \wedge \hat{\be_{k_r}} \wedge \cdots
\wedge \hat{\be_{k_1}} \wedge \cdots
\\
&\quad
 =
(-1)^B \det (G)
\be_{1} \wedge \be_{2} \wedge \cdots,
\end{align*}
where
$$
G
 =
\begin{pmatrix}
\left( X_{\alpha_i,j} \right)_{1 \le i \le r, 1 \le j \le s}
&
\left( X_{\alpha_i,k_{r+1-j}} \right)_{1 \le i, j \le r}
\\
\left( X_{-\delta_{s+1-i}-1,j} \right)_{1 \le i, j \le s}
&
\left( X_{-\delta_{s+1-i}-1,k_{r+1-j}} \right)_{1 \le i \le s, 1 \le j \le r}
\end{pmatrix},
$$
and
$$
B = \sum_{j=1}^r (k_j-1-s-(r-j)).
$$
Since $k_j = \# \{ m \in M : m \ge -\beta_j-1 \} = s + \beta_j+1-n_j$, we have
$$
B
 =
\sum_{j=1}^r \beta_j - \sum_{j=1}^r n_j - \binom{r}{2}.
$$
Since $\sum_{i=1}^r n_i = \# \{ (i,j) : \beta_i > \delta_j \}$ 
and $\sum_{j=1}^s n'_j = \# \{ (i,j) : \beta_i < \delta_j \}$, 
we have $\sum_{j=1}^r n_j + \sum_{j=1}^s n'_j = sr$.
Hence  we have
\begin{equation}
\label{eq:xi1}
\xi_\lambda^U
 =
(-1)^{\sum_{j=1}^r \beta_j + \sum_{j=1}^s \delta_j + \binom{r}{2} + \binom{s}{2} + rs}
\det (G).
\end{equation}

Since Proposition~\ref{prop:frame_entry} gives us
$$
G
 =
\begin{pmatrix}
\left( (-1)^{j-1} {\tilde u}^{(j)}_{\alpha_i} \right)_{1 \le i \le r, 1 \le j \le s}
&
\left( (-1)^{\beta_{r+1-j}} {\tilde z}_{\alpha_i,\beta_{r+1-j}} \right)_{1 \le i, j \le r}
\\
O
&
\left( (-1)^{\beta_{r+1-j}-\delta_{s+1-i}-(s+1-i)} {\tilde v}_{\beta_{r+1-j}}^{(s+1-i)} \right)_{1 \le i \le s, 1 \le j \le r}
\end{pmatrix},
$$
we have
\begin{align}
&
\det G
\notag
\\
&\quad
=
(-1)^{ \sum_{j=1}^s \delta_j + \binom{s+1}{2} + \binom{s}{2} + \sum_{j=1}^r \beta_j }
\det
\begin{pmatrix}
\left( {\tilde u}^{(j)}_{\alpha_i} \right)_{1 \le i \le r, 1 \le j \le s}
&
\left( {\tilde z}_{\alpha_i,\beta_{r+1-j}} \right)_{1 \le i, j \le r}
\\
O
&
\left( {\tilde v}_{\beta_{r+1-j}}^{(s+1-i)} \right)_{1 \le i \le s, 1 \le j \le r}
\end{pmatrix}
\notag
\\
&\quad
=
(-1)^{ s + \sum_{j=1}^r \beta_j + \sum_{j=1}^s \delta_j }
\cdot
(-1)^{rs + \binom{r}{2} + \binom{s}{2} }
\det
\begin{pmatrix}
\left( {\tilde z}_{\alpha_i,\beta_j} \right)_{1 \le i, j \le r}
&
\left( {\tilde u}^{(j)}_{\alpha_i} \right)_{1 \le i \le r, 1 \le j \le s}
\\
\left( {\tilde v}_{\beta_j}^{(i)} \right)_{1 \le i \le s, 1 \le j \le r}
&
O
\end{pmatrix}
\label{eq:xi2}
\end{align}
by permuting rows and columns.
Therefore we obtain (\ref{eq:Pcoord_det}) under the assumption (\ref{eq:cond_generic}) 
by combining (\ref{eq:xi1}) and (\ref{eq:xi2}).

\begin{remark}
If $\lambda$ and $\mu$ do not satisfy the condition (\ref{eq:cond_generic}), 
then the degree of the determinant becomes smaller than $r+s$. 
We omit details of computation, because Proposition~\ref{prop:Pcoord_det} guarantees that 
(\ref{eq:Pcoord_det}) is valid for such cases.
\end{remark}

\section{
Generalized Giambelli identity for skew Schur functions
}

For two partitions $\lambda$ and $\mu$, the skew Schur function $s_{\lambda/\mu}$ is defined by
\begin{equation}
\label{eq:skewSchur}
s_{\lambda/\mu} = \sum_\nu c^\lambda_{\mu,\nu} s_\nu,
\end{equation}
where $c^\lambda_{\mu,\nu}$ is the Littlewood--Richardson coefficient given by
\begin{equation}
\label{eq:LR}
s_\mu s_\nu = \sum_\lambda c^\lambda_{\mu,\nu} s_\lambda.
\end{equation}
It is known that
$s_{\mu/\mu} = 1$ and $s_{\lambda/\mu} = 0$ unless $\lambda \supset \mu$.

\begin{prop}
\label{prop:KP_sol}
For a fixed partition $\mu$, we define a formal power series $\tau^\mu(x)$ by putting
$$
\tau^\mu(x) = \sum_\lambda s_{\lambda/\mu}(y) s_\lambda(x),
$$
where $\lambda$ runs over all partitions and $y$ is another set of variables.
Then $\tau^\mu(x)$ is a solution to the KP hierarchy satisfying the condition (\ref{eq:cond1}) 
and (\ref{eq:cond2}).
\end{prop}

\begin{demo}{Proof}
By using (\ref{eq:skewSchur}), (\ref{eq:LR}) and the Cauchy identity (see \cite[I.4. (4.1)]{Mac}), we have
$$
\tau^\mu(x)
 =
\sum_\lambda \sum_\nu c^\lambda_{\mu,\nu} s_\nu(y) s_\lambda(x)
 =
\sum_\nu s_\nu(y) s_\mu(x) s_\nu(x)
 =
s_\mu(x) \exp \left( \sum_{n=1}^\infty n x_n y_n \right).
$$
It follows from the Pl\"ucker relation (Proposition~\ref{prop:KP_Plucker}) that 
the Schur function $s_\mu(x)$ is a solution of the KP hierarchy \cite{SS}.
Also, by using (\ref{eq:KP}), we see that, if $\tau(x)$ is a solution of the KP hierarchy, 
then so is $\tau(x) \exp \left( \sum_{n \ge 1} c_n x_n \right)$ for any constants $\{ c_n \}_{n \ge 1}$.
Therefore $\tau^\mu(x)$ is a solution of the KP hierarchy.
\end{demo}

If we apply Theorem~\ref{thm:main} to the special solution given in Proposition~\ref{prop:KP_sol}, 
then we can derive a generalization of the Giambelli identity to skew Schur functions as follows.
First we recall the Lascoux--Pragacz formula for skew Schur functions.
If $\lambda = (\alpha_1, \cdots, \alpha_r | \beta_1, \cdots, \beta_r)$ 
and $\mu = (\gamma_1, \cdots, \gamma_s | \delta_1, \cdots, \delta_s)$ are partitions in the Frobenius notation, 
then we have
\begin{equation}
\label{eq:LP}
s_{\lambda/\mu}
 =
(-1)^s
\det \begin{pmatrix}
 \left(
  s_{(\alpha_i|\beta_j)}
 \right)_{1 \le i, j \le r}
&
 \left( h_{\alpha_i - \gamma_j} \right)_{1 \le i \le r, 1 \le j \le s}
\\
 \left( e_{\beta_j - \delta_i} \right)_{1 \le i \le s, 1 \le j \le r}
&
 O
\end{pmatrix},
\end{equation}
where $h_k$ and $e_k$ are the $k$th complete and elementary symmetric functions respectively.
For any pair of sequences $\alpha = (\alpha_1, \dots, \alpha_r)$ and $\beta = (\beta_1, \dots, \beta_r)$ 
of nonnegative integers, we define $s_{(\alpha_1, \dots, \alpha_r|\beta_1, \dots, \beta_r)/\mu}$ by 
the same formula as (\ref{eq:LP}):
$$
s_{(\alpha|\beta)/\mu}
 =
(-1)^s
\det \begin{pmatrix}
 \left(
  s_{(\alpha_i|\beta_j)}
 \right)_{1 \le i, j \le r}
&
 \left( h_{\alpha_i - \gamma_j} \right)_{1 \le i \le r, 1 \le j \le s}
\\
 \left( e_{\beta_j - \delta_i} \right)_{1 \le i \le s, 1 \le j \le r}
&
 O
\end{pmatrix}.
$$
If the entries of $\alpha$ (or $\beta$) are not distinct, then $s_{(\alpha|\beta)/\mu} = 0$.
Otherwise, if $\sigma$ and $\tau \in \Sym_r$ be permutations such that
$\alpha_{\sigma(1)} > \cdots > \alpha_{\sigma(r)}$ and $\beta_{\tau(1)} > \cdots > \beta_{\tau(r)}$, 
then we have $s_{(\alpha|\beta)/\mu} = \sgn(\sigma \tau) s_{\lambda/\mu}$, 
where $\lambda$ is a partition given 
by the Frobenius notation $(\alpha_{\sigma(1)}, \cdots, \alpha_{\sigma(r)} | \beta_{\tau(1)}, \cdots, \beta_{\tau(r)})$.
Then we have the following skew-generalization of Giambelli identity (\ref{eq:Schur_Giambelli}). 
See \cite{O} for a more direct proof by using a generalization of the Sylvester formula for determinants.

\begin{corollary}
\label{cor:Schur_skewGiambelli}
For two partitions $\lambda = (\alpha_1, \cdots, \alpha_r|\beta_1, \cdots, \beta_r)$ and 
$\mu = (\gamma_1, \cdots, \gamma_s|\delta_1, \cdots, \delta_s)$, we have
\begin{multline}
\label{eq:skewGiambelli}
s_{\lambda/\mu}
\\
 =
(-1)^s
\det
\begin{pmatrix}
 \left( s_{(\alpha_i,\gamma_1, \cdots, \gamma_s|\beta_j,\delta_1, \cdots, \delta_s)/\mu} \right)_{1 \le i, j \le r}
&
 \left(
  s_{(\alpha_i, \gamma_1, \cdots, \widehat{\gamma_j}, \cdots, \gamma_s|\delta_1, \cdots, \delta_s)/\mu}
 \right)_{1 \le i \le r, 1 \le j \le s}
\\
 \left(
  s_{(\gamma_1, \cdots, \gamma_s|\beta_j, \delta_1, \cdots, \widehat{\delta_i}, \cdots, \delta_s)/\mu}
 \right)_{1 \le i \le s, 1 \le j \le r}
&
 O
\end{pmatrix}.
\end{multline}
\end{corollary}

If $\mu = \emptyset$, then (\ref{eq:skewGiambelli}) reduces to the Giambelli identity (\ref{eq:Schur_Giambelli}).


\end{document}